\documentclass[12pt, draftcls,onecolumn]{IEEEtran}

\usepackage[english]{babel}
\usepackage[utf8]{inputenc}

\usepackage{cite} 
\usepackage{xspace}
\usepackage{filecontents}

\usepackage{float}
\usepackage{pgf} 

\usepackage{graphicx}
\usepackage{tikz}
\usepackage[cmex10]{amsmath}

\usepackage{bm}

\usepackage{amssymb} 
\usepackage{array}

\usepackage{subfigure}
\usepackage{caption2}

\usepackage{amsfonts}
\usepackage{amsthm}   

\usepackage{enumerate}

\usepackage{extarrows}

\newtheorem{theorem}{Theorem}

\newtheorem{remark}{Remark}

\usepackage{enumerate}

\usepackage{extarrows}

\newcommand{\cmse}{\mathsf{CMSE}}
\newcommand{\ccmse}{\widetilde{\mathsf{CMSE}}}

\newcommand{\diag}{\mathsf{diag}}

\begin{document}

\title{Over-the-Air Computation with Spatial-and-Temporal Correlated Signals}
\author{Wanchun Liu, Xin Zang, Branka Vucetic and Yonghui Li
\vspace{-0.5cm}
}

\maketitle

\begin{abstract}
\let\thefootnote\relax\footnote{The authors are with School of Electrical and Information Engineering, the University of Sydney, Australia.
	Emails:	\{wanchun.liu,\ xin.zang,\ branka.vucetic,\ yonghui.li\}@sydney.edu.au. (\emph{Yonghui Li is the corresponding author.})
}Over-the-air computation (AirComp) leveraging the superposition property of wireless multiple-access channel (MAC),
is a promising technique for effective data collection and computation of large-scale wireless sensor measurements in Internet of Things applications.
Most existing work on AirComp only considered computation of spatial-and-temporal independent sensor signals, though in practice different sensor measurement signals are usually correlated.
In this letter, we propose an AirComp system with spatial-and-temporal correlated sensor signals, and formulate the optimal AirComp policy design problem for achieving the minimum computation mean-squared error (MSE).
We develop the optimal AirComp policy with the minimum computation MSE in each time step by utilizing the current and the previously received signals.
We also propose and optimize a low-complexity AirComp policy in closed form with the performance approaching to the optimal policy. 
\end{abstract}

\begin{IEEEkeywords}
	Over-the-air computing, wireless sensor networks, multiple-access channel, spacial-and-temporal correlation.
\end{IEEEkeywords}

\section{Introduction}
In the past few years, over-the-air computation (AirComp) has emerged as a promising solution for large-scale wireless sensor data fusion in Internet of Things applications~\cite{Goldenbaumletter,liu2019over,GuangxuMIMO,Wen19,Gunduz,consensus}.
In general, an AirComp system consists of multiple sensors and a receiver to perform data fusion by computing a pre-determined function of the sensors' measurement signals.
Each sensor of the AirComp system sends its \emph{pre-processed} measurement signal simultaneously to the receiver over a multiple-access channel (MAC). By leveraging  the \emph{superposition} property of MAC and applying a \emph{post-processing} function to the received sum (superimposed) signal, the receiver obtains the desired function output of the sensors' signals (see e.g.~\cite{liu2019over} for details).

From a computation theoretic perspective,  pre-processing and post-processing function design problems of AirComp systems for different fusion functions have been comprehensively investigated in \cite{GoldenbaumTCOM,GoldenbaumWIOPT,Katabi}. 
From a wireless communications perspective, 
due to the imperfection of practical MAC with non-zero
receiver noise, unequal channel coefficients and limited transmit power budgets, the output signal of a MAC is not a perfect sum of the pre-processed sensor signals as expected. 
Thus, most recent work focuses on optimal AirComp transmitter and receiver policy design for achieving the minimum computation mean-squared error (MSE) of the sum of pre-processed signals~\cite{Goldenbaumletter,liu2019over,GuangxuMIMO,Wen19}.
In \cite{Goldenbaumletter}, the computation MSE of the sum signal under imperfect channel state information was investigated.
In \cite{liu2019over}, 
the optimal single-antenna AirComp design and the scaling law analysis of  computation MSE and transmit power consumption were derived in terms of the number of sensors.
In~\cite{GuangxuMIMO} and~\cite{Wen19}, the transmitting and receiving beamforming of multi-antenna AirComp systems were considered to minimize the computation MSE.
In particular, AirComp has been applied to emerging wireless applications such as wireless distributed machine learning~\cite{Gunduz}, and over-the-air consensus~\cite{consensus}.

%
%

Most existing work on AirComp system assumed spatial-and-temporal independent sensor signals for ease of analysis and optimization~\cite{Goldenbaumletter,liu2019over,GuangxuMIMO,Wen19}.
However, in practice, different sensor measurements are correlated and the current measurements are normally related to the previous ones.
Spatial-and-temporal sensor signals are commonly considered in the literature of remote state estimation~(see e.g. \cite{KangTWC,AlexControl}), but have not been considered in AirComp systems to the best of our knowledge.
In particular, unlike the conventional time-independent AirComp system, where only the current received signal is used to estimate the sum signal of the sensors, 
the time correlation introduces a new challenge in AirComp system design: how to effectively utilize both the current and the sequence of previously received signals for optimal computation of the sum of current sensor signals?

In this paper, we focus on AirComp system design with spatial-and-temporal correlated sensor signals. The main contributions are summarized as follows.
\begin{itemize}
	\item We propose an AirComp system with spatial-and-temporal correlated sensor signals for the first time in the literature, and formulate the optimal AirComp policy design problem for achieving the minimum computation MSE.
	\item We derive the optimal AirComp policy to achieve the minimum computation MSE in each time step by utilizing the current and the previously received signals.
	\item We propose a low-complexity AirComp policy, where the receiver stores the most recent $(l+1)$ received signals and applies a linear filter to it as the computation output. The optimal AirComp filter is derived in closed form. Once an AirComp filter is given, we show that the optimal transmit power of each sensor can be obtained by solving a classic quadratically constrained quadratic program (QCQP) problem. An alternating minimization approach is adopted to jointly design the AirComp filter and the transmit power.
\end{itemize}

\emph{Notations}: $\{x\}^j_i$ denotes the sequence $\{x_i,x_{i+1},\dots,x_{j}\}$.
$\rho(\mathbf{A})$ is the spectral radius of a square matrix $\mathbf{A}$, i.e., the largest absolute value of its eigenvalues. $\diag\{\mathbf{v}\}$ denotes a diagonal matrix generated by the vector $\mathbf{v}$. $\mathbf{1}$ is a all-one column vector. $\mathbf{0}_{i,j}$ is a $i\times j$ all-zero matrix. $[\mathbf{A}]_{i,j}$ is the element in the $i$th row and $i$th column of $\mathbf{A}$.

\section{AirComp System with Spatial-and-Temporal Correlated Signals}
We consider a single-antenna AirComp system with $K$ sensors and a receiver.
\subsection{Spatial-and-Temporal Correlated Signals}
Given a sampling period $T$, sensor $k$'s signal at time slot~$t$ is $x_{k,t}\in \mathbb{R},\forall k\in \mathcal{K}, t \in \mathbb{N}_0$, where $\mathcal{K}\triangleq\{1,\cdots,K\}$.
We assume that the sensors' signals are spatial-and-temporal correlated, and are modeled as a Gauss-Markov process~\cite{kailath2000linear}:
\begin{equation}\label{eq:x}
\mathbf{x}_{t+1} = \mathbf{A} \mathbf{x}_{t} + \mathbf{w}_t, t\in\mathbb{N}_0
\end{equation}
where $\mathbf{x}_{t} \triangleq [x_{1,t},x_{2,t},\dots,x_{K,t}]^\top \in \mathbb{R}^K$, $\mathbf{A}\in\mathbb{R}^{K\times K}$ is the process parameter, and $\mathbf{w}_t \triangleq [w_{1,t},w_{2,t},\dots,w_{K,t}]^\top  \in\mathbb{R}^K$ is a white Gaussian noise with zero mean and a covariance matrix $\mathbf{V}_w$.
Thus, each sensor's current signal depends on its and other sensors' previous signals.
Further, it is assumed that $\rho(\mathbf{A})<1$, i.e., $\mathbf{x}_t$ has a steady state distribution with zero mean and a (bounded) converged steady-state covariance matrix $\mathbf{V}_x \triangleq \mathsf{E}[\mathbf{x}_t \mathbf{x}^\top_t]$.
From~\cite{kailath2000linear}, $\mathbf{V}_x$ of the Gauss-Markov process~\eqref{eq:x} is the unique positive semi-definite solution of the discrete Lyapunov matrix equation below
\begin{equation}\label{eq:lyap}
\mathbf{A}\mathbf{V}_x\mathbf{A}^\top-\mathbf{V}_x+\mathbf{V}_w=\mathbf{0}.
\end{equation}
In the scenario that $\rho(\mathbf{A})\geq 1$, $\mathbf{x}_t$ has an unbounded covariance matrix when $t\rightarrow \infty$, which is beyond the scope of the paper.  

\subsection{AirComp System and Computation MSE}
Each sensor linearly scales its signal by its Tx-scaling factor $b_k$, and then sends $b_k x_{k,t}$ to the receiver simultaneously over a MAC at time slot $t$.
Let $h_{k}$ denote the channel coefficient between sensor $k$ and the receiver.
We focus on the scenario that the sensor signals change much faster than the channel status. Thus, $h_k$ is assumed to be time-invariant~\cite{LiuJIoT}.
The received signal at time slot $t$ is~\cite{liu2019over}
\begin{equation}\label{signal}
y_t= \sum_{k=1}^{K} h_k b_k x_{k,t} + z_t, t \in \mathbb{N}_0,
\end{equation}
where $z_t$ is the receiver-side additive white Gaussian noise (AWGN) with zero mean and variance $\sigma^2_z$.
For ease of notation but without loss of generality, we use $b_k$ to denote $h_k b_k$ in the rest of the analysis.
Let $\mathbf{b}\triangleq[b_1,\dots,b_K]^\top$ denote the Tx-scaling vector. Then, \eqref{signal} can be rewritten as
\begin{equation}
y_t = \mathbf{b}^\top \mathbf{x} + z_t.
\end{equation}
Note that similar to the previous work~\cite{liu2019over,Zang20}, we focus on a fundamental real channel scenario, i.e., $h_k,b_k,y_t,z_t \in \mathbb{R},\forall k \in \mathcal{K}$,
though the analytical framework and results in the paper can be readily
extended to the complex channel scenario by representing $y_t$ and $z_t$ as two-dimensional vectors, and $\mathbf{b}$ as a $K\times 2$ matrix with real and imaginary parts.

Different from the scenario with time-independent signals (see e.g.,~\cite{GuangxuMIMO}), the previously received signals $\{y\}^{t-1}_0$ are correlated with and hence contain information of the current signal $\mathbf{x}_t$.
Thus, it is assumed that the previous received signals are stored at the receiver.

The AirComp system with time-correlated signals in~\eqref{eq:x} is designed to calculate the sum of the sensors' signals $\sum_{k=1}^{K} x_{k,t}$ in each time slot based on the current and previously received signals $\{y\}^t_0$.
Let $\chi_t$ denote the calculated sum signal.
Then, the computation distortion of the AirComp system at time slot $t$ is measured by the computation MSE defined as
\begin{equation}\label{mse_1}
\cmse_t \triangleq \mathsf{E}\left[\lvert \chi_t -\sum_{k=1}^{K} x_{k,t} \rvert^2 \Big\vert \{y\}^t_0 \right], \forall t \in \mathbb{N}.
\end{equation}

In the following, we will investigate the optimal AirComp policy $\pi(\cdot)$ to compute $\chi_t = \pi(\{y\}^t_0)$ with the smallest computation MSE.

\section{Optimal AirComp Policy Design}
We introduce a virtual vector $\mathbf{\tilde{x}}_t \triangleq [\tilde{x}_{1,t},\tilde{x}_{1,t},\dots,\tilde{x}_{K,t}]^\top \in  \mathbb{R}^K$.
Although an AirComp policy is to find a scalar as the computation output, without loss of generality, the policy can be treated as to find a vector $\mathbf{\tilde{x}}_t$ in each time slot and calculate  the sum of it, i.e., $\chi_t \triangleq  \sum_{k=1}^{K} \tilde{x}_{k,t}$. 
To find the optimal $\chi_t$ achieving the minimum computation MSE, one needs to find the optimal virtual vector $\mathbf{\tilde{x}}_t$.
From \eqref{mse_1}, the optimal AirComp policy design problem is formulated as
\begin{equation}\label{prob:1}
\min_{\mathbf{\tilde{x}}_t}\quad \cmse_t \triangleq \mathsf{E}\left[\lvert \sum_{k=1}^{K} \tilde{x}_{k,t} -\sum_{k=1}^{K} x_{k,t} \rvert^2 \Big\vert \{y\}^t_0 \right], \forall t \in \mathbb{N}.
\end{equation}

The target function in problem \eqref{prob:1} can be rewritten as
\begin{align}
\label{eq:mse}
& \cmse_t = \mathsf{E}\left[\mathbf{1}^\top (\mathbf{\tilde{x}}_t - \mathbf{x}_t) (\mathbf{\tilde{x}}_t - \mathbf{x}_t)^\top \mathbf{1} \Big\vert \{y\}^t_0 \right]\\
\notag
&=\mathsf{E}\left[\mathbf{1}^\top \mathbf{\tilde{x}}_t \mathbf{\tilde{x}}_t^\top  \mathbf{1}
 - 2\times \mathbf{1}^\top \mathbf{\tilde{x}}_t \mathbf{{x}}_t^\top  \mathbf{1}
 + \mathbf{1}^\top \mathbf{{x}}_t \mathbf{{x}}_t^\top  \mathbf{1}
  \Big\vert \{y\}^t_0 \right]\\
\notag
&=\!\!\int_{\mathbb{R}^K}\!\!\!\!\left(\mathbf{1}^\top \mathbf{\tilde{x}}_t \mathbf{\tilde{x}}_t^\top  \mathbf{1}
- 2\times\! \mathbf{1}^\top \mathbf{\tilde{x}}_t \mathbf{{x}}_t^\top  \mathbf{1}
+ \mathbf{1}^\top \mathbf{{x}}_t \mathbf{{x}}_t^\top  \mathbf{1} \right)\!\! f\!\!\left(\mathbf{x}_t \vert \{y\}^t_0\right)\!\! \mathrm{d} \mathbf{x}_t\\
\notag
&= \mathbf{1}^\top \mathbf{\tilde{x}}_t \mathbf{\tilde{x}}_t^\top  \mathbf{1}\!+\!\! \int_{\mathbb{R}^K}\!\!\!\!\left(
- 2\times\! \mathbf{1}^\top \mathbf{\tilde{x}}_t \mathbf{{x}}_t^\top  \mathbf{1}
\!+\! \mathbf{1}^\top \mathbf{{x}}_t \mathbf{{x}}_t^\top  \mathbf{1} \right) \!\!f\!\!\left(\mathbf{x}_t \vert \{y\}^t_0\right)\!\! \mathrm{d} \mathbf{x}_t,
\end{align}
where $f\left(\mathbf{x}_t \vert \{y\}^t_0\right)$ is the conditional probability density function of $\mathbf{x}_t$.

It can be shown that $\cmse_t$ is a convex function of $\mathbf{\tilde{x}}_t$ as the Hessian matrix of $\cmse_t$ is $2 \times \mathbf{1}\mathbf{1}^\top$, which is positive semi-definite.
Since any local minimum of a convex function is also a global minimum~\cite{boyd2004convex}, we only need to find a local minimum of $\cmse_t$.
Then, by letting the derivative equal to zero, we have
\begin{equation}\label{eq:derivative}
\frac{\partial \cmse_t}{\partial \mathbf{\tilde{x}}_t} = 2 \times \mathbf{1}\mathbf{1}^\top \mathbf{\tilde{x}}_t - 2 \times \mathbf{1}\mathbf{1}^\top \int_{\mathbb{R}^K} \mathbf{x}_t f\left(\mathbf{x}_t \vert \{y\}^t_0\right) \mathrm{d} \mathbf{x}_t = \mathbf{0}.
\end{equation}
Although \eqref{eq:derivative} has many solutions, it is clear that 
\begin{equation}\label{eq:opti_virtual}
\mathbf{\tilde{x}}_t = \mathbf{\hat{x}}_t \triangleq \int_{\mathbb{R}^K} \mathbf{x}_t f\left(\mathbf{x}_t \vert \{y\}^t_0\right) \mathrm{d} \mathbf{x}_t
\end{equation}
is a solution achieving the minimum $\cmse_t$. 
Note that $\mathbf{\hat{x}}_t$ in \eqref{eq:opti_virtual} is the conditional expectation of the current signal $\mathbf{x}_t$.
Based on the linear estimation theory~\cite{kailath2000linear}, $\mathbf{\hat{x}}_t$ is obtained by the classic Kalman filter (KF) in an iterative prediction-correction way as below:
\begin{subequations} \label{eq:kalman}
	\begin{align}\label{eq:kalman_1}
&\textbf{Prediction: } \hat{\mathbf{x}}_{t \vert t-1}=\mathbf{A}\hat{\mathbf{x}}_{t-1},\\
\label{eq:kalman_1_1}
&\textbf{Prediction MSE Matrix: } \!\mathbf{M}_{t \vert t-1}\!\!=\!\mathbf{A}\mathbf{M}_{t-1}\mathbf{A}^\top\!\!\!+\!\mathbf{V}_w,\\
\label{eq:kalman_1_2}
&\textbf{Kalman Gain Vector: } \mathbf{K}_t\!=\!{\mathbf{M}_{t|t-1}\mathbf{b}}\left({\sigma_z^2+\!\mathbf{b}^\top\mathbf{M}_{t|t-1}\mathbf{b}}\right)^{\!-1}\!,\\
\label{eq:kalman_1_3}
&\textbf{Correction: } \hat{\mathbf{x}}_{t}=\hat{\mathbf{x}}_{t \vert t-1}+\mathbf{K}_t(y_t-\mathbf{b}^\top\hat{\mathbf{x}}_{t \vert t-1}),\\
\label{eq:kalman_2}
&\textbf{Estimation MSE matrix: }  \mathbf{M}_{t}\!=\!(\mathbf{I}\!-\!\mathbf{K}_t\mathbf{b}^\top\!)\mathbf{M}_{t \vert t-1}.
	\end{align}
\end{subequations}
In particular, $\mathbf{M}_t$ is the estimation error covariance matrix of $\mathbf{\hat{x}}_t$, i.e.,
\begin{equation}
\mathbf{M}_t \triangleq \mathsf{E}\left[(\mathbf{\hat{x}}_t-\mathbf{x}_t) (\mathbf{\hat{x}}_t-\mathbf{x}_t)^\top \Big\vert \{y\}^t_0 \right].
\end{equation}
Taking \eqref{eq:opti_virtual} into \eqref{eq:mse}, the minimum $\cmse_t$ is equal to $\mathbf{1}^\top \mathbf{M}_t \mathbf{1}$.
Therefore, we have the following result.
\begin{theorem}\label{theo:1}
	\normalfont
Given the Tx-scaling factor vector $\mathbf{b}$, the optimal AirComp policy for achieving the minimum computation MSE is to first apply the KF  defined in \eqref{eq:kalman} to estimate the sensor signal vector $\mathbf{x}_t$ and then calculate the sum of the estimate $\mathbf{\hat{x}}_t$. The minimum computation MSE is $\mathbf{1}^\top \mathbf{M}_t \mathbf{1}$, where $\mathbf{M}_t$ is give in \eqref{eq:kalman_2}.
\end{theorem}

\begin{remark}
It is non-trivial that the KF-based state estimation method leads to the optimal computation of the sum of the original signals.
The KF is optimally designed to achieve the minimum state estimation MSE, i.e., $\mathsf{E}\left[(\mathbf{\hat{x}}_t-\mathbf{x}_t)^\top (\mathbf{\hat{x}}_t-\mathbf{x}_t)\right] = \mathsf{Tr}\left(\mathbf{M}_t\right)$~\cite{kailath2000linear}, which is not equal to the design target of the AirComp problem~\eqref{prob:1}, i.e., $\mathsf{E}\left[\mathbf{1}^\top (\mathbf{\hat{x}}_t-\mathbf{x}_t) (\mathbf{\hat{x}}_t-\mathbf{x}_t)^\top \mathbf{1}\right] = \mathbf{1}^\top\mathbf{M}_t \mathbf{1}$. Thus, without the rigorous proof of Theorem~\ref{theo:1}, one cannot tell whether the KF-based estimation-then-sum method is the optimal or not.

Also, it is interesting to see that in the AirComp system with time correlated signals, one needs to recover (estimate) each sensor's signal first based on the previous and current observations (received signals) and then calculate the sum signal; while in the conventional time independent scenario~\cite{liu2019over,Zang20}, the optimal policy is to estimate the sum signal directly rather than the original signals.
\end{remark}

Taking \eqref{eq:kalman_1_1} into \eqref{eq:kalman_1_2} and then into \eqref{eq:kalman_1_3}, it is clear that the computation of $\mathbf{\hat{x}}_t$ requires $K\times K$ matrix multiplications. So the computation complexity of the optimal AirComp policy in each time step is $\mathcal{O}(K^3)$ based on the big O notation, 
when the total number of sensors $K$ is very large.
Further, since $\mathbf{M}_t$ is a complex and implicit function of the Tx-scaling factor $\mathbf{b}$, it is difficult to design the optimal $\mathbf{b}$ for achieving the minimum computation MSE, $\mathbf{1}^\top \mathbf{M}_t \mathbf{1}$.

In the following, we will propose and optimize a low complexity AirComp policy, based on which an optimal Tx-scaling factor can be obtained.

\section{Low-Complexity AirComp Policy and Transmission Power Control}
\subsection{Low-Complexity AirComp Policy}
We propose a low-complexity AirComp policy: the receiver stores the most recent $(l+1)$ received signals 
\begin{equation}\label{eq:y}
 \mathbf{y}_t \triangleq [y_{t-l},y_{t-l+1}, \dots, y_{t}]^\top,\ t\geq l, l \geq 1,
\end{equation}
and applies a linear filter $\mathbf{g} \triangleq [g_0,g_1,\dots,g_l]^\top \in \mathbb{R}^{l+1}$ to it as the computation output $\chi_t$, i.e.,
\begin{equation} \label{eq:filter}
\chi_t = \mathbf{g}^\top \mathbf{y}_t.
\end{equation}
Intuitively, a longer filter utilizing more previous received signals achieves a better computation MSE.
Compared with the optimal AirComp policy in Theorem~\ref{theo:1} with third-degree polynomial complexity, the policy~\eqref{eq:filter} has a linear computation complexity of $\mathcal{O}(l)$ and is invariant with the total number of sensors.

The computation MSE of the low-complexity policy~is
\begin{equation} \label{eq:mse_low}
\ccmse_t=\mathsf{E}\left[\vert \mathbf{g}^\top\mathbf{y}_t-\mathbf{1}^\top\mathbf{x}_t\vert^2\right].
\end{equation}
Using the Gauss-Markov property of $\{\mathbf{x}_t\}$ in~\eqref{eq:x}, the $i$th previously received signal can be written as
\begin{equation} \label{eq:y_individual}
y_{t-i}  = \mathbf{b}^\top \mathbf{A}^i \mathbf{x}_{t-l} + c_{i,t}, i\in\{0,1,\dots,l\}
\end{equation}
where
\begin{equation}
c_{i,t} = 
z_{t-l+i}+\sum_{m=1}^{i}\mathbf{b}^\top\mathbf{A}^{m-1}\mathbf{w}_{t-l+m-1}.
\end{equation}
Taking \eqref{eq:y_individual} into \eqref{eq:y}, we have
\begin{equation}\label{eq:y_t}
\mathbf{y}_t=\mathbf{Mx}_{t-l}+\mathbf{c}_t,
\end{equation}
where $\mathbf{c}_t \triangleq [c_{0,t}, c_{2,t},\dots,c_{l,t}]^\top$, and 
\begin{equation}\label{eq:M}
\mathbf{M} \triangleq 
\left[(\mathbf{A}^0)^\top \mathbf{b},
(\mathbf{A}^1)^\top \mathbf{b},\dots,
(\mathbf{A}^l)^\top \mathbf{b}\right]^\top.
\end{equation}
Further, $\mathbf{x}_t$ can be represented as a function of $\mathbf{x}_{t-l}$ and the noise terms as
\begin{equation} \label{eq:x_t}
\mathbf{x}_t=\mathbf{A}^l\mathbf{x}_{t-l}+\sum_{i=1}^{l}\mathbf{A}^{l-i}\mathbf{w}_{t-l+i-1}.
\end{equation}
Taking \eqref{eq:y_t} and \eqref{eq:x_t} into \eqref{eq:mse_low}, it can be obtained that
\begin{equation}\label{eq:mse_2}
\begin{aligned}
&\ccmse_t\\
&=\mathsf{E}\!\!\left[\!\Big\vert \mathbf{g}^\top\!(\mathbf{Mx}_{t-l}+\mathbf{c}_t) \!- \!\mathbf{1}^\top\!(\mathbf{A}^l\mathbf{x}_{t-l}\!+\!\sum_{i=1}^{l}\!\!\mathbf{A}^{l-i}\mathbf{w}_{t-l+i-1})\Big\vert^2\!\right].
\end{aligned}
\end{equation}

To find the optimal low-complexity AirComp filter $\mathbf{g}$, we have the following optimization problem:
\begin{equation} \label{prob:2}
 \min_{\mathbf{g}} \qquad \eqref{eq:mse_2}.
\end{equation}
By introducing and deriving the covariance matrices $\mathbf{V}_c \triangleq \mathsf{E}\left[\mathbf{c}_t\mathbf{c}_t^\top\right]$ and $\mathbf{C}_{i} \triangleq \mathsf{E}\left[\mathbf{c}_t\mathbf{w}_{t-i}^\top\right], i\in \{1,2,\dots,l\}$,  with the properties that $\{z_t\}$ and $\{\mathbf{w}_t\}$ are independent white Gaussian processes, the target function can be simplified as
\begin{equation} \label{msec2}
\begin{aligned}
&\ccmse_t
=\mathbf{g}^\top\mathbf{M}\mathbf{V}_{x}\mathbf{M}^\top\mathbf{g}\!-\!\mathbf{g}^\top\mathbf{M}\mathbf{V}_{x}(\mathbf{A}^l)^\top\mathbf{1}\!-\!\mathbf{1}^\top\mathbf{A}^l\mathbf{V}_{x}\mathbf{M}^\top\mathbf{g}\\
&+\mathbf{1}^\top\mathbf{A}^l\mathbf{V}_{x}(\mathbf{A}^l)^\top\mathbf{1}+\mathbf{g}^\top\mathbf{V}_c\mathbf{g}+\sum_{i=1}^{l}\mathbf{1}^\top\mathbf{A}^{l-i}\mathbf{V}_{w}(\mathbf{A}^{l-i})^\top\mathbf{1}\\
&-\mathbf{g}^\top\sum_{i=1}^{l}\mathbf{C}_{l+1-i}(\mathbf{A}^{l-i})^\top\mathbf{1}-\mathbf{1}^\top\sum_{i=1}^{l}\mathbf{A}^{l-i}\mathbf{C}_{l+1-i}^\top\mathbf{g},
\end{aligned}
\end{equation}
which is time independent. Thus, we drop the time index in $\ccmse_t$.
It is clear that $\ccmse$ has a quadratic form and is strictly convex, and the optimal $\mathbf{g}$ can be obtained directly by solving the linear equation:
\begin{equation}
\begin{split}\label{eq:V_c}
&\frac{\partial \ccmse }{\partial \mathbf{g}}\!=\!\mathbf{0} \!=\! 2\mathbf{g}^\top\mathbf{M}\mathbf{V}_{x}\mathbf{M}^\top\!-\!\mathbf{1}^\top\mathbf{A}^l\mathbf{V}_{x}^\top\mathbf{M}^\top\!-\!\mathbf{1}^\top\mathbf{A}^l\mathbf{V}_{x}^\top\mathbf{M}^\top\\&+2\mathbf{g}^\top\mathbf{V}_c-\mathbf{1}^\top\sum_{i=1}^{l}\mathbf{A}^{l-i}\mathbf{C}_{l+1-i}^\top-\mathbf{1}^\top\sum_{i=1}^{l}\mathbf{A}^{l-i}\mathbf{C}_{l+1-i}^\top.
\end{split}	
\end{equation}
Then, we have the following result.
\begin{theorem}
	\normalfont
Given the Tx-scaling factor $\mathbf{b}$, the optimal length-$(l+1)$ linear filter $\mathbf{g}$ of the low-complexity AirComp policy~is 
\begin{equation}\label{optimalg}
\mathbf{g}^\top\!=\!\!\left(\!\mathbf{1}^\top\mathbf{A}^l\mathbf{V}_{x}^\top\mathbf{M}^\top\!\!+\!\mathbf{1}^\top\!\sum_{i=1}^{l}\mathbf{A}^{l-i}\mathbf{C}_{l+1-i}^\top\!\right)\!\!(\mathbf{M}\mathbf{V}_{x}\mathbf{M}^\top\!+\!\mathbf{V}_c)^{-1},
\end{equation}
where
\vspace{-0.2cm}
\begin{equation} \label{eq:Vc}
       	\mathbf{V}_c= \diag\{\underbrace{\sigma^2_z,\dots,\sigma^2_z}_{l+1}\}+ \mathbf{\tilde{V}}_c,
\end{equation}
 \begin{equation}
 [\mathbf{\tilde{V}}_c]_{i,j} = \left\lbrace 
 \begin{aligned}
 &0, && i=0 \text{ or } j=0,\\
 & \mathbf{b}^\top \mathbf{U}_{i,j} \mathbf{b}, && 1 \leq i \leq j \\
 &v_{j,i}, && i>j \geq 1,
 \end{aligned}
 \right.
 \end{equation} 
 \begin{equation}
\mathbf{U}_{i,j} = \sum_{u=1}^{i} \mathbf{A}^{i-u} \mathbf{V}_w (\mathbf{A}^{j-u})^\top, 1 \leq i \leq j,
 \end{equation}
and
\vspace{-0.2cm}
\begin{equation}\label{eq:C}
\mathbf{C}_{i}=\begin{bmatrix}
\mathbf{0}_{(l+1-i) \times K}\\
\mathbf{b}^\top \mathbf{A}^0\mathbf{V}_{w}\\
\mathbf{b}^\top \mathbf{A}^1\mathbf{V}_{w}\\
\vdots\\
\mathbf{b}^\top\mathbf{A}^{i-1}\mathbf{V}_{w}
\end{bmatrix}.
\end{equation}
 
\end{theorem} 
 
\subsection{Tx-Scaling Factor Design}
Given a low-complexity AirComp policy~\eqref{eq:filter}, we will optimize the Tx-scaling factor $\mathbf{b}$ under a power constraint to minimize the computation MSE in~\eqref{msec2}.
Although the computation MSE in the current form is an implicit function of $\mathbf{b}$, from the definitions of $\mathbf{M}$, $\mathbf{V}_c$ and $\mathbf{C}_i$ in \eqref{eq:M}, \eqref{eq:V_c} and \eqref{eq:C}, respectively, it is not difficult to see that \eqref{msec2} is a quadratic function of $\mathbf{b}$.
In the following, we will simplify each term of \eqref{msec2}  with respect to $\mathbf{b}$.

For the first term, we have
\begin{equation} \label{eq:detail_1}
\begin{aligned}
\mathbf{g}^\top\mathbf{M}\mathbf{V}_{x}\mathbf{M}^\top\mathbf{g}&= \sum_{i=0}^{l}\sum_{j=0}^{l} g(i) g(j) \mathbf{b}^\top \mathbf{A}^i \mathbf{V}_x (\mathbf{A}^j)^\top \mathbf{b}\\
&= \mathbf{b}^\top \left(\sum_{i=0}^{l}\sum_{j=0}^{l} \mathbf{G}_i \mathbf{A}^i \mathbf{V}_x (\mathbf{A}^j)^\top \mathbf{G}_j\right) \mathbf{b},
\end{aligned}
\end{equation}
where $\mathbf{G}_{i}$ denotes $\diag\{\underbrace{g_i, \dots, g_i}_{K}\}, i \in \{0,1,\dots,l\}$.
Similarly, it is easy to have
\begin{equation}\label{eq:detail_2}
\begin{aligned}
\mathbf{g}^\top\mathbf{M}\mathbf{V}_{x}(\mathbf{A}^l)^\top\mathbf{1}
=\mathbf{b}^\top\sum_{i=0}^{l}\mathbf{G}_{i}\mathbf{A}^i\mathbf{V}_{x}(\mathbf{A}^l)^\top\mathbf{1},
\end{aligned}	
\end{equation} 
and
\vspace{-0.2cm}
 \begin{equation}\label{eq:detail_3}
 \begin{split}
 \mathbf{g}^\top\mathbf{V}_c\mathbf{g}=\sigma_{z}^2\mathbf{g}^\top\mathbf{g}
 +\mathbf{b}^\top \left(\sum_{i=1}^{l}\sum_{j=1}^{l} \mathbf{G}_i \mathbf{U}'_{i,j} \mathbf{G}_j\right)
 \mathbf{b},
 \end{split}	
 \end{equation}
where $\mathbf{U}'_{i,j} = \mathbf{U}_{i,j}$ and $\mathbf{U}_{j,i}$ for cases $i\leq j$ and $i>j$, respectively.
Further, from the definition of \eqref{eq:C}, it can be obtained that
\begin{equation} \label{eq:detail_4}
\begin{aligned}
\mathbf{g}^\top\sum_{i=1}^{l}\mathbf{C}_{l+1-i}(\mathbf{A}^{l-i})^\top\mathbf{1}
\!=\!\mathbf{b}^\top\!\sum_{i=1}^{l}\sum_{j=i}^{l}\mathbf{G}_{j}\mathbf{A}^{j-i} \mathbf{V}_w (\mathbf{A}^{l-i})^\top\mathbf{1}.
\end{aligned}
\end{equation}
Taking \eqref{eq:detail_1}-\eqref{eq:detail_4} into \eqref{msec2}, we have
\begin{equation}\label{eq:mse_b}
	\begin{aligned}
	&\ccmse	=\mathbf{b}^\top\left(\sum_{i=0}^{l}\sum_{j=0}^{l} \mathbf{G}_i \mathbf{A}^i \mathbf{V}_x (\mathbf{A}^j)^\top \mathbf{G}_j\right)\mathbf{b}\\
	&-2\mathbf{b}^\top\!\!\left(\!\sum_{i=0}^{l}\!\mathbf{G}_{i}\mathbf{A}^i\mathbf{V}_{x}(\mathbf{A}^l)^\top\mathbf{1} \!+\!\sum_{i=1}^{l}\sum_{j=i}^{l}\!\mathbf{G}_{j}\mathbf{A}^{j-i} \mathbf{V}_w (\mathbf{A}^{l-i})^\top\mathbf{1}\!\right)\\
	&+\mathbf{1}^\top\mathbf{A}^l\mathbf{V}_{x}(\mathbf{A}^l)^\top\mathbf{1}+\sigma_{z}^2\mathbf{g}^\top\mathbf{g}+\sum_{i=1}^{l}\mathbf{1}^\top\mathbf{A}^{l-i}\mathbf{V}_{w}(\mathbf{A}^{l-i})^\top\mathbf{1}.
	\end{aligned}
\end{equation}

We assume that each sensor has an average transmit power limit $P_k, \forall k \in \mathcal{K}$. Then, the average transmit power constraint is $
\mathsf{E}\left[ \vert b_k x_{k,t} \vert^2 \right] \leq h^2_k P_k,
$ with a quadratic form as
\begin{equation}\label{eq:power_con}
\mathbf{b}^\top \mathbf{P}_k \mathbf{b} \leq P_k/\sigma^2_{x,k}, k \in \mathcal{K},
\end{equation}
where $\sigma^2_{x,k} \triangleq \mathsf{E}[x^2_{k,t}]$, which is the $k$th diagonal element of the covariance matrix $\mathbf{V}_x$, and $\mathbf{P}_k$ is an all-zero matrix except for the $k$th diagonal element, which equals to one.

Considering the power constraint~\eqref{eq:power_con}, the optimal Tx-scaling factor design problem is formulated as
\begin{equation}\label{prob:3}
\min_{\mathbf{b}} 	\qquad \eqref{eq:mse_b}, \qquad 
	\text{subject to } \eqref{eq:power_con}.
\end{equation}
\eqref{prob:3} is a classic QCQP problem, which can be effectively solved using semi-definite programming (SDP)~\cite{Luo}. 

\subsection{Joint Design of AirComp Policy and Tx-Scaling Factor}\label{sec:joint}
The computation MSE in \eqref{eq:mse_b} is a non-convex function of the AirComp filter $\mathbf{g}$ and the Tx-scaling factor $\mathbf{b}$.
To the best of the our knowledge,
there is no general approach to find the optimal $\mathbf{g}$ and $\mathbf{b}$ jointly. Therefore, we adopt the \emph{alternating minimization approach} by alternately
solving problems \eqref{prob:2} and \eqref{prob:3} for $\mathbf{g}$ and $\mathbf{b}$, respectively, while fixing the other.
Note that alternating minimization is a widely applicable and empirically successful approach for optimization problems with different subsets of variables~\cite{Yu16}.
The effectiveness of the approach in the AirComp joint design problem will be verified via numerical results in the following section.

%
%
%
%
%
%
%
%
%
\section{Numerical Results}
In this section, we numerically evaluate the computation MSE of the optimal and low-complexity policies with different number of sensors $K$.
Unless otherwise stated, we set $\mathbf{A}=\alpha \mathbf{I}$~\cite{AlexControl}, where $0<\alpha<1$, $\mathbf{V}_x=\mathbf{I}$, $\sigma^2_z=1$ and $P_k =10,\forall k\in\mathcal{K}$~\cite{Zang20}. A large $\alpha$ means that the process ${\mathbf{x}_t}$ is tightly time-correlated. From \eqref{eq:lyap}, we have $\mathbf{V}_w = (1-\alpha^2) \mathbf{I}$.
We assume normalized Rayleigh fading channels for evaluating the AirComp system performance that averages over $10^5$ random channel realizations.

\begin{figure}[t]	
	\centering
	\includegraphics[scale=0.8]{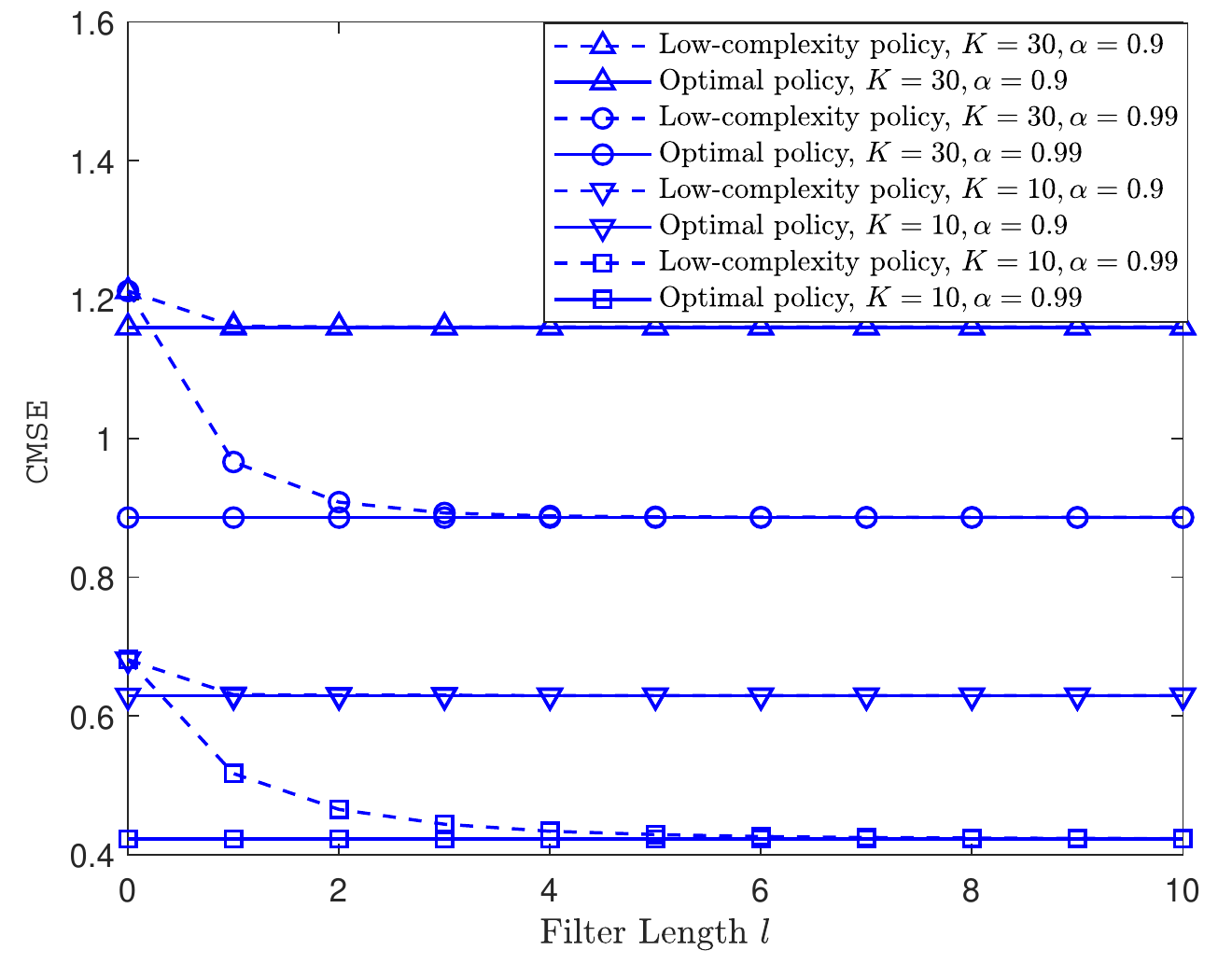}
	\caption{The computation MSE of the optimal and low-complexity AirComp policies.}
	\label{fig:performance}
	\vspace{-0.5cm}
\end{figure}

In Fig.~\ref{fig:performance}, we first plot the computation MSE of low-complexity policy with AirComp filter $\mathbf{g}$ and Tx-scaling factor $\mathbf{b}$ achieved by the alternating minimization approach in Section~\ref{sec:joint} with $50$ iteration rounds. 
We see that the computation MSE decreases with the length of the filter $(l+1)$ as expected, and a larger sensor number $K$ leads to a higher commutation MSE.
It is interesting to see that $l=1$ has achieved the optimal computation MSE when $\alpha = 0.9$, while it is larger than $5$ when $\alpha = 0.99$. So we need a longer filter to achieve the minimum computation MSE when the process ${\mathbf{x}_t}$ is tightly time-correlated.
Then, we plot the computation MSE achieved by the optimal AirComp policy in Theorem~\ref{theo:1} and the optimal Tx-scaling factor of the low-complexity policy mentioned earlier.
We see that the performance of the low-complexity policy approaches to the optimal one when $l$ is large.

\begin{figure}[t]	
	\centering	
	\includegraphics[scale=0.8]{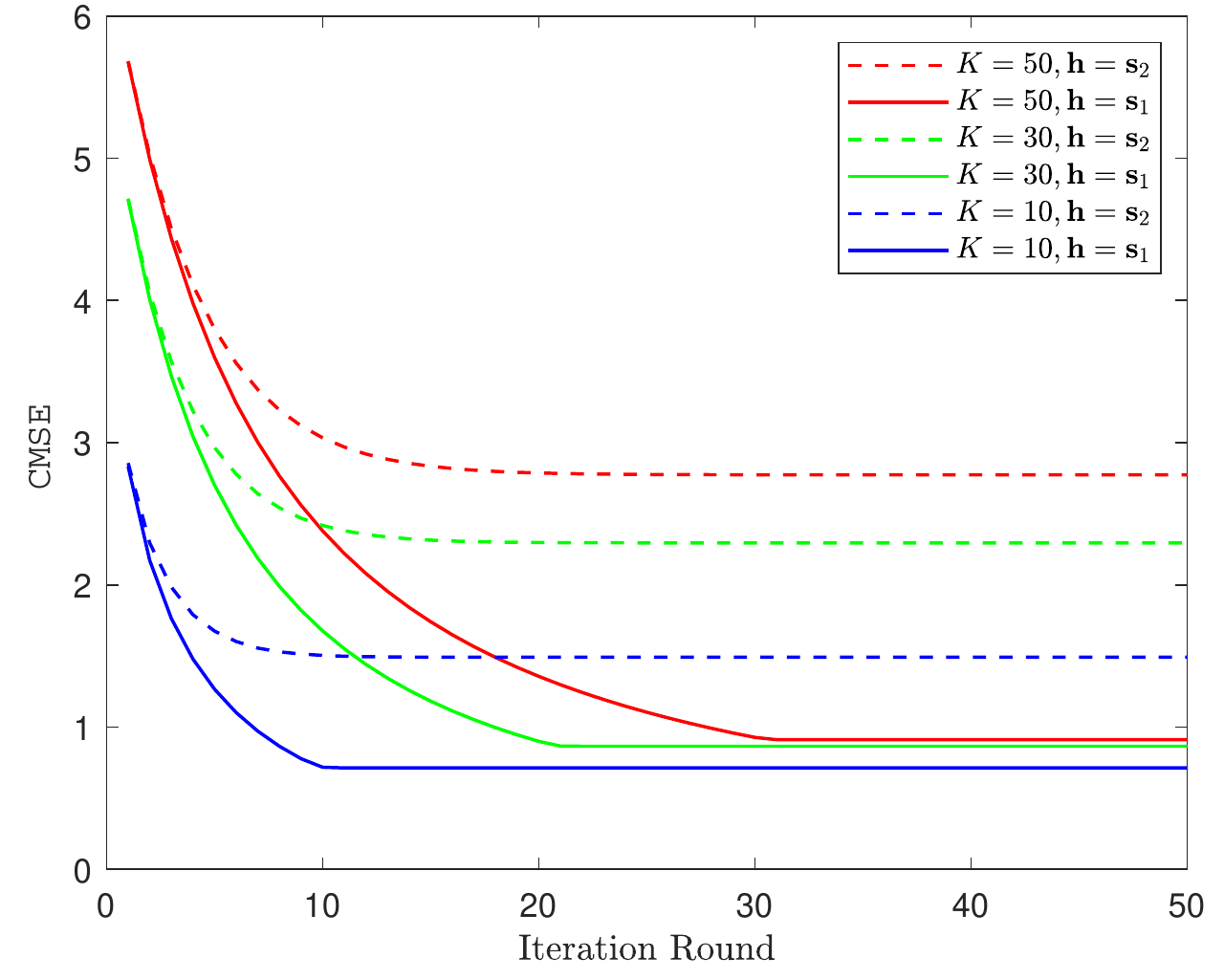}
	\caption{The computation MSE of the low-complexity AirComp policy with different iteration rounds.}	
	\label{fig:convergence}
	\vspace{-0.5cm}	
\end{figure}

In Fig.~\ref{fig:convergence}, we evaluate the convergence speed of the alternating minimization approach of the low-complexity AirComp policy with $l=4$ under an equal channel power gain $\mathbf{h} = \mathbf{s}_1=[1,1,\cdots,1]^\top$ of different sensors and an unequal channel power gain $\mathbf{h} = \mathbf{s}_2$, where $\mathbf{s}_2$ is generated by $K$ evenly spaced values between $0.1$ and $1.9$.
We see that the proposed algorithm converges within $30$ iterations in different scenarios.
It can be observed that a larger number of sensors and a higher divergence of channel conditions of different channels require more iteration rounds to converge.


\section{Conclusions}
In this letter, we have 
proposed an AirComp system with spatial-and-temporal correlated sensor signals, and derived the optimal AirComp policy to achieve the minimum computation MSE in each time step by utilizing the current and the previously received signals.
We have also proposed and optimized a low-complexity AirComp policy with a linear AirComp filter, based on which the optimal transmit power of each sensor has been optimized.
An alternating minimization approach has been adopted to jointly design the AirComp filter and the Tx-scaling factor.


\bibliographystyle{IEEEtran}
\end{document}